# The impact of learning assistants on inequities in physics student outcomes

Ben Van Dusen[1], Jada-Simone S. White[1], and Edward A. Roualdes[2]

[1]*California State University Chico, Department of Science Education*
*101 Holt Hall, Chico, CA, 95929, USA*

[2]*California State University Chico, Department of Mathematics and Statistics*
*204 Holt Hall, Chico, CA, 95929, USA*

This study investigates how Learning Assistants (LAs) and related course features are associated with inequities in student learning in introductory university physics courses. 2,868 physics students' paired pre- and post-test scores on concept inventories from 67 classes in 16 LA Alliance member institutions are examined in this investigation. The concept inventories included the Force Concept Inventory, Force and Motion Conceptual Evaluation, and the Conceptual Survey of Electricity and Magnetism. Our analyses include a multiple linear regression model that examines the impact of student (e.g. gender and race) and course level variables (e.g. presence of LAs and Concept Inventory used) on student learning outcomes (Cohen's d effect size) across classroom contexts. The presence of LAs was found to either remove or invert the traditional learning gaps between students from dominant and non-dominant populations. Significant differences in student performance were also found across the concept inventories.

## I. INTRODUCTION

The existence of entrenched disparities in student performances across gender, racial, and ethnic groups has been well documented [1]. These "achievement gaps" [2] have been the focus of many calls for reform in the STEM disciplines to better meet the needs of of students from non-dominant communities [3]. The National Research Council report examining the state of Discipline Based Education Research [1] states that while, "DBER clearly indicates that student-centered instructional strategies can positively influence students' learning… Most of the studies the committee reviewed were not designed to examine differences in terms of gender, ethnicity, socioeconomic status, or other student characteristics." [pg. 136-137] The NRC goes on to identify examining the performance of students from non-dominant cultures as an important direction for future research.

The Learning Assistant (LA) model was developed for several reasons, including to improve undergraduate STEM student learning outcomes by increasing faculty use of research-based instructional strategies in undergraduate courses [4]. Since the introduction of the first LA workshop in 2007, the number of institutions with LA programs has grown from 3 to over 90 institutions [5]. In response to this growth, a coalition of LA using institutions (LA Alliance) was created. Each of the 90 institutions in the LA Alliance has its own contextual affordances and constraints that act to shape the ways it implements its LA model. Even within a given institution, variation in classroom contexts, such as what discipline they are teaching, can lead instructors to use LAs in significantly different ways. The creation of the LA Alliance made it possible to collect data across institutional settings using the LA Supported Student Outcomes (LASSO) online assessment tool (see methods section for details). The LASSO dataset has been used to document the broad trends in student outcomes in LA supported courses [6]. In this paper we examine associations between LA-supported classroom features and achievement gaps in physics courses.

## II. RESEARCH QUESTIONS

By examining student outcomes, demographics, and classroom features we investigate the questions: (1) What impacts do LAs have on learning gaps in physics, if any? (2) What impacts do concept inventories have on student learning gaps in physics, if any?

## III. CONCEPTUAL FRAMEWORK

Critical Race Theory (CRT) provides a framework for operationalizing race and racism in learning environments [7]. A central tenet of CRT is that racism is deeply ingrained in our social fabric in a way that allows its endemic nature to go largely unacknowledged and unexamined. Ladson-Billings & Tate [7] propose that, "class- and gender-based explanations are not powerful enough to explain all of the difference (or variance) in school experiences and performance" and that race, "continues to be significant in explaining inequity in the United States" [pg. 51]. A second tenet of CRT is the importance of giving voice to members of marginalized groups. Creating space for a minority student to tell their reality supports the "psychic preservation of marginalized groups" while "catalyzing the necessary cognitive conflict to jar dysconscious racism" [pg 57-58]. A third tenet of CRT is that the interests of marginalized groups are primarily advanced when they align with the interests of those with power. Milner explained that, "quite often, those in power are not interested in having to negotiate or question their own privilege to provide opportunities to empower people of color or to 'level the playing field'" [8, pg. 391]. These three tenets of CRT suggest that racism is endemic to classes and that, to advance racial equity, it is critical that we leverage students' voices and experiences in service of learning.

The LA model is grounded in the idea that student learning is facilitated by engagement with peers on group-worthy tasks [9]. The nature of these group-worthy tasks can vary across class context, but they often involve some form of argumentation in which students express and defend their ideas to their peers. These type of activities can give students voice in the classroom by shifting the power structures such that authority is distributed amongst the students through their use of evidence. These classroom structures are aligned with those of the scientific community in which authority does not come from an individual's rank or title, but rather in evidence from nature [10]. These "interactive" engagement activities have been associated with significant improvement in student outcomes [11]. Because LA-supported activities align with the goals of physics instructors and can give students voice, there is reason to believe that they may create a sustainable decrease in classroom inequities.

Cultural-Historical Activity Theory (CHAT) [12] provides us a lens to examine the roles that features of a learning environment play in exacerbating or ameliorating student inequities. CHAT emerged from the works of Vygotsky [13] and his student Leont'ev [14]. Vygotsky and Leont'ev were instrumental in blending Marxist ideas with educational research in what became known as socioculturalism [12]. In a radical departure from cognitive psychology, sociocultural perspectives of cognition and learning broadened the unit of analysis from the human brain to include the social and physical environments in which an activity is embedded [15]. CHAT proposes that there are seven social and material components of an activity system that interact dynamically to produce an outcome (Fig. 1). Typical components of the activity systems in our study include: (1) Subject – physics student; (2) Object – a concept inventory; (3) Rules – the classroom and cultural norms of behavior; (4) Community – students, LAs, and the teacher; (5) Divisions of labor – students engage in groups, LAs support groups, and the instructor oversees the activity; (6) Mediating Artifacts – whiteboards, clickers, PowerPoint slides, carts, etc.; and (7) Outcomes – pre & post scores on a concept inventory [16]. The interactive nature of these components is often visualized [12]. It is assumed that these activity systems are dynamic in nature. In a process that is analogous to changing the value of a resistor in a complex circuit, changing any individual component in the activity system can significantly alter the interactions between the other components of the system. CHAT directs our analyses to focus on the seven components of the classroom system.

Nasir and Hand [17] utilize a sociocultural perspective to argue that the underachievement of minority students is the product of both a multilevel process involving both micro-processes (e.g. individual student interactions) and macro-structures (e.g. political climate). To understand the role that these processes have on students Nasir and Hand recommend the development of models that examine the interaction of student and setting level factors.

By transforming the components of the classroom system, including the rules, communities, and divisions of labor, LAs have the potential to create classroom environments in which the interests of students from non-dominant populations and faculty converge. Specifically, classes can be transformed in ways that create space for marginalized students to voice and utilize their lived-experiences in service of learning physics content. Through the development of a multiple linear regression model, we examine how the classroom social and physical structures are interacting to perpetuate or ameliorate physics classroom inequities.

## IV. METHODS

### A. Data Collection

Data for this investigation were collected using the LA Supported Student Outcomes (LASSO) online assessment tool. LASSO is a free tool that is hosted on the LA Alliance website [18] and allows faculty (LA-using or not) to easily administer Research-Based Assessment Instruments as pre and post tests to their students online. To use LASSO, faculty provide course-level information, select their assessment(s), and upload a list of student names and emails. When faculty launch an assessment their students receive emails with unique links to complete their pre-tests online. As part of completing the instrument, students answer a set of demographic questions. The LASSO system allows faculty to track their students' participation and send reminder emails. At the end of the semester students receive another set of emails with unique links to their post-tests. Faculty can download their students' responses as well as a summary report that shows the distribution of their students' pre and post scores, normalized learning gains (Hake score), and effect sizes (Cohen's d). As of the Fall 2016 semester LASSO is hosting 15 instruments across the STEM disciplines.

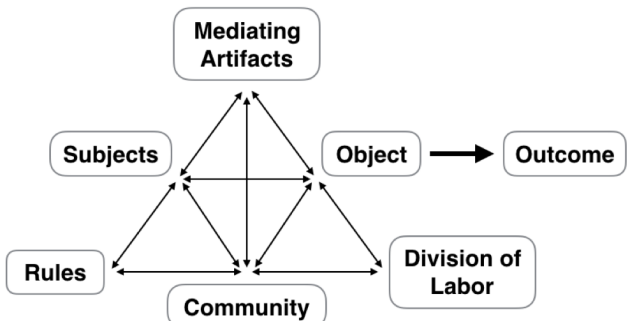

**FIG 1.** The seven dynamically interactive components of an activity system.

TABLE I. Cleaned data counts by instrument.

| Instrument | Institutions | Courses | Students (% Non-dom.) |
|---|---|---|---|
| FCI | 9 | 31 | 1,005 (41%) |
| FMCE | 8 | 15 | 1,109 (73%) |
| CSEM | 2 | 21 | 754 (45%) |
| Total | 16 | 67 | 2,868 (55%) |

TABLE II. Cleaned data counts by LA presence.

| | | FCI | FMCE | CSEM | Total |
|---|---|---|---|---|---|
| No LAs | Majority | 363 | 27 | 0 | 390 |
| | Non-dom. | 221 | 51 | 0 | 272 |
| LAs | Majority | 230 | 271 | 413 | 914 |
| | Non-dom. | 191 | 760 | 341 | 1292 |

In this investigation we examined data from courses that used the Force Concept Inventory (FCI) [19], Force and Motion Conceptual Evaluation (FMCE) [20], and Conceptual Survey of Electricity and Magnetism (CSEM) [21]. Over the first three semesters of data collection, prior to data cleaning, a total of 6,190 unique student responses were collected from 124 courses at 16 institutions on the three instruments in these analyses.

## V. DATA ANALYSIS

Data cleaning involved removing student data for any of the following reasons: (1) Less than 80% of the concept inventory questions completed, (2) no matching pre or post test, (3) incomplete demographic data, (4) outliers that may not have followed instructions or cheated (d<-2 or >4), (4) less than 10 matched data sets in a course (either due to low enrollment or participation). Once student results were cleaned, there were 2,868 usable pre-post pairs of responses from 67 courses in 16 institutions (Table I). Based on historical physics classroom demographics [3], The 1,304 White or Asian, non-Hispanic/Latino, male students were classified as culturally "dominant" while the other 1,564 students were classified as culturally "non-dominant" (Table II). Paired responses were assigned a Cohen's d effect size. Cohen's d is a measure of change (in this case from pre to post scores) in units of standard deviations [22].

To examine the impact of LAs on classroom inequities, the learning gap is defined as the non-dominant students' mean effect size minus the dominant students' mean effect size. The learning gap is examined with and without LAs for each concept inventory (Fig. 2). Paired t-tests with Bonferroni corrections were used to test for statistical significance.

A multiple linear regression model was developed to measure potential effects of the concept inventories on learning gaps. Because only 20% of our data was from classes without LAs, they were not included in the model. The model tested student dominance status, the concept inventories, and their interaction effects in LA-supported courses. Model assumptions of normality and homoscedasticity were checked visually and no obvious aberrations were found.

## VI. FINDINGS

Examining the learning gap across the assessments indicates that LAs are associated with improved outcomes for non-dominant students (Fig. 2). The learning gap was significantly negative (i.e. dominant students outperformed their non-dominant peers) in courses without LAs (Fig. 2a: $t_{597.6}=3.67$; $p<0.005$). The learning gap was significantly positive (i.e. non-dominant students outperformed dominant students), however, in courses with LAs where ($t_{2068.2}=-2.63$; $p<0.025$). The same trend was seen on the FCI (fig 2b), where the learning gap was significantly negative without LAs ($t_{470.0}=2.64$; $p<0.025$) but significantly positive in classes with LAs ($t_{397.6}=-2.76$; $p<0.025$). The FMCE showed a similar trend (Fig. 2c). While there was not a significant learning gap for courses without LAs, it was

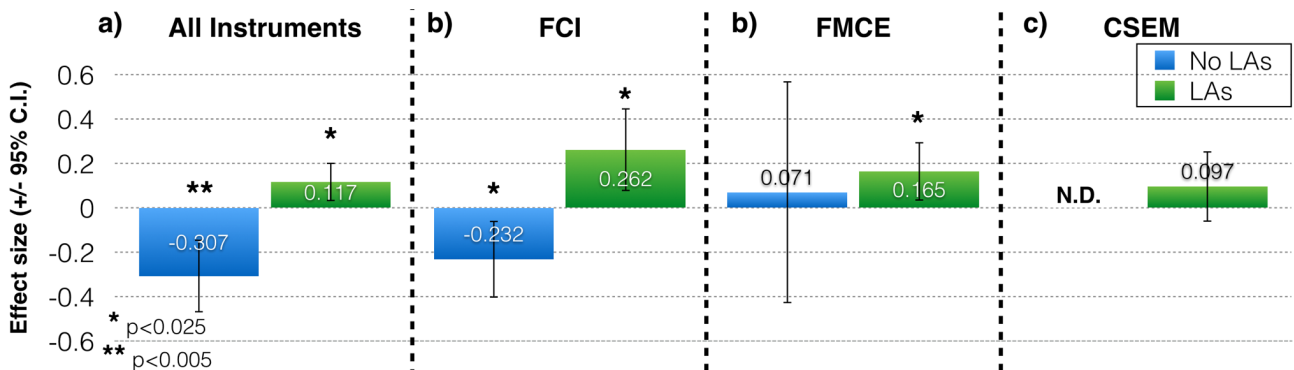

**FIG 2.** Learning gap for students with and without LAs (mean effect size of non-dominant students – mean effect size of dominant students).

TABLE III. Effects of non-dominant status and concept inventory on student effect size in LA supported classes.

|  | Coefficient | S.E. | t value | p |
|---|---|---|---|---|
| (Intercept) | 1.066*** | 0.068 | 15.589 | 0.000 |
| **Non-dom.** | **0.262**** | 0.102 | 2.579 | 0.010 |
| **FMCE** | **-0.200*** | 0.093 | -2.151 | 0.032 |
| CSEM | -0.074 | 0.085 | -0.866 | 0.387 |
| Non-dom.*FMCE | -0.097 | 0.125 | -0.776 | 0.438 |
| Non-dom.*CSEM | -0.165 | 0.127 | -1.304 | 0.193 |

Sig. Codes: *** p<0.001; ** p<0.01; * p<0.05

significantly positive in LA-supported courses ($t_{556.7}$=-2.45; p<0.025). We had no CSEM data for courses without LAs, but the courses with LAs did not find a significant learning gap (Fig. 2d).

Our multiple linear regression model set the intercept to dominant students who took the FCI. The attribute coefficients are measures of their differences from the intercept. To predict the mean effect size for any given student, start with the intercept value (1.066) and add the coefficients for any attributes that match the student of interest. Interaction effect coefficients are added if both statuses are true (e.g. a non-dominant student taking the CSEM would add -0.165).

The model identified the coefficients for non-dominant status and the FMCE (shown in bold) to both be statistically significant. The non-dominant coefficient indicates that in courses with LAs non-dominant students are projected to have mean effect sizes that are 0.262 higher than their dominant peers. The FMCE coefficient indicates that in courses with LAs the mean student effect sizes on the FMCE is projected to be 0.2 lower than on the FCI.

## VII. DISCUSSION

Our analyses indicate that variation in classroom contexts (i.e. the activity system) are associated with significant variations in physics student inequities. Changing a classroom's "Community" and "Division of Labor" to accommodate LAs appear to at erase or invert classroom inequities in learning "Outcomes". For courses with LAs, non-dominant students had "Outcomes" that were better than (FCI, FMCE) or equivalent to (CSEM) their dominant peers. This is a remarkable finding given the historical persistence of these learning gaps. Identifying the cause of these shifts in learning gaps will require additional investigation. It is reasonable to hypothesize that it is partially driven by LAs' transformation of the interactions between the "Rules" and "Community" of classroom systems. In LA-supported classes, students are often encouraged to voice their experiences in making sense of physical phenomenon.

The improved "Outcomes" of non-dominant students in LA supported courses was seen again in our model (Table II). Our model also showed significant differences across "Objects" (i.e. concept inventories) in LA-supported activity systems. Our model indicates that student effect sizes may be partially driven by the concept inventory used in the course.

## VIII. CONCLUSION & FUTURE WORK

This investigation provides an initial examination of the impact of LAs on inequity in physics classes. The findings are promising in that the presence of LAs is strongly associated with the removal of traditional learning gaps. Additional analyses are required to understand the nature of the roles that LAs play in ameliorating learning gaps. Future investigations will disaggregate effects across non-dominant populations. This work was funded in part by NSF-IUSE Grant No. DUE-1525338 and is Contribution No. LAA-034 of the International Learning Assistant Alliance.